\documentclass[letter]{aa} 

\usepackage{graphicx}
\usepackage{txfonts}
\usepackage{url}
\usepackage{natbib}
\bibpunct{(}{)}{;}{a}{}{,}
\usepackage{enumitem}
\usepackage{array}
\usepackage{comment}

\begin{document} 

   \title{New jet feature in the parsec-scale jet of the blazar OJ\,287 connected to the 2017 teraelectronvolt flaring activity}

  \author{R.~Lico\inst{1,2,3}\fnmsep\thanks{Email: rlico@iaa.es}, C.~Casadio\inst{4,5,2}, S.~G.~Jorstad\inst{6,7}, J.~L.~G\'omez\inst{1}, A.~P.~Marscher\inst{6}, E.~Traianou\inst{1,2}, J.-Y.~Kim\inst{8,2}, G.-Y.~Zhao\inst{1}, A.~Fuentes\inst{1}, I.~Cho\inst{1}, T.~P.~Krichbaum\inst{2}, O.~Hervet\inst{9}, S.~O'Brien\inst{10}, B.~Boccardi\inst{2}, I.~Myserlis\inst{11,2}, I.~Agudo\inst{1}, A.~Alberdi\inst{1}, Z.~R.~Weaver\inst{6}, J.~A.~Zensus\inst{2}} 

   \institute{Instituto de Astrof\'{\i}sica de Andaluc\'{\i}a-CSIC, Glorieta de la Astronom\'{\i}a s/n, 18008 Granada, Spain.
   \and Max-Planck-Institut f\"{u}r Radioastronomie, Auf dem H\"{u}gel 69, D-53121 Bonn, Germany.
   \and INAF Istituto di Radioastronomia, via Gobetti 101, 40129 Bologna, Italy.
   \and Foundation for Research and Technology - Hellas, IESL \& Institute of Astrophysics, Voutes, 7110 Heraklion, Greece.
   \and Department of Physics, University of Crete, GR-70013 Heraklion, Greece.
   \and Institute for Astrophysical Research, Boston University, 725 Commonwealth Avenue, Boston, MA 02215.
   \and Astronomical Institute, St. Petersburg State University, Universitetskij Pr. 28, Petrodvorets, St. Petersburg 198504, Russia.
   \and Korea Astronomy and Space Science Institute, Daedeok-daero 776, Yuseong-gu, Daejeon 34055, Republic of Korea.
   \and Santa Cruz Institute for Particle Physics and Department of Physics, University of California, Santa Cruz, CA 95064, USA.
   \and Physics Department, McGill University, Montreal, QC H3A 2T8, Canada.
   \and Institut de Radioastronomie Millim\'etriquea, Avenida Divina Pastora 7, Local 20, E-18012 Granada, Spain. 
              }

   \date{Received ...; accepted ...}

  \abstract
   {In February 2017 the blazar OJ\,287, one of the best super-massive binary-black-hole-system candidates, was detected for the first time at very high energies (VHEs; E$>100$\,GeV) with the ground-based $\gamma$-ray observatory VERITAS.}  
   {Very high energy $\gamma$ rays are thought to be produced in the near vicinity of the central engine in active galactic nuclei. For this reason, and with the main goal of providing useful information for the characterization of the physical mechanisms connected with the observed teraelectronvolt flaring event, we investigate the parsec-scale source properties by means of high-resolution very long baseline interferometry observations.}
   {We use 86\,GHz Global Millimeter-VLBI Array (GMVA) observations from 2015 to 2017 and combine them with additional multiwavelength radio observations at different frequencies from other monitoring programs. We investigate the source structure by modeling the brightness distribution with two-dimensional Gaussian components in the visibility plane.}
   {In the GMVA epoch following the source VHE activity, we find a new jet feature (labeled K) at $\sim0.2$\,mas from the core region and located in between two quasi-stationary components (labeled S1 and S2). Multiple periods of enhanced activity are detected at different radio frequencies before and during the VHE flaring state.}
   {Based on the findings of this work, we identify as a possible trigger for the VHE flaring emission during the early months of 2017 the passage of a new jet feature through a recollimation shock (represented by the model-fit component S1) in a region of the jet located at a de-projected distance of $\sim10$ pc from the radio core.}

\authorrunning{R. Lico et al.}
\titlerunning{Parsec-scale jet properties of the blazar OJ\,287 during the 2017 TeV flaring activity.}

   \maketitle


\section{Introduction} \label{introduction}
Super-massive binary black-hole (SMBBH) systems are commonly believed to form during hierarchical galaxy formation as an outcome of merger events \citep[e.g.,][]{Begelman1980}. 
The blazar OJ\,287 ($z = 0.306$) is currently considered one of the best SMBBH system candidates, supported by a number of lines of observational evidence and theoretical arguments \citep{Lehto1996,Valtonen2008}.
While the vast majority of blazar objects show erratic variability across the electromagnetic spectrum, from the optical light curve of OJ\,287 two quasi-periodic variability patterns have been revealed, on timescales of $\sim12$ yr and $\sim60$ yr, respectively \citep{Valtonen2006}.  
Within the context of a gravitationally bound system of two black holes, the $\sim12$ yr and $\sim60$ yr optical outbursts are connected with the orbital period and the periastron advance timescales, respectively. 
Additional scenarios, not necessarily requiring the presence of a secondary black hole, are also being investigated with the goal of explaining the optical emission variability \citep[e.g.,\ ][]{Britzen2018, Liska2018}.

\begin{figure*}
\begin{center}
\includegraphics[scale=0.5]{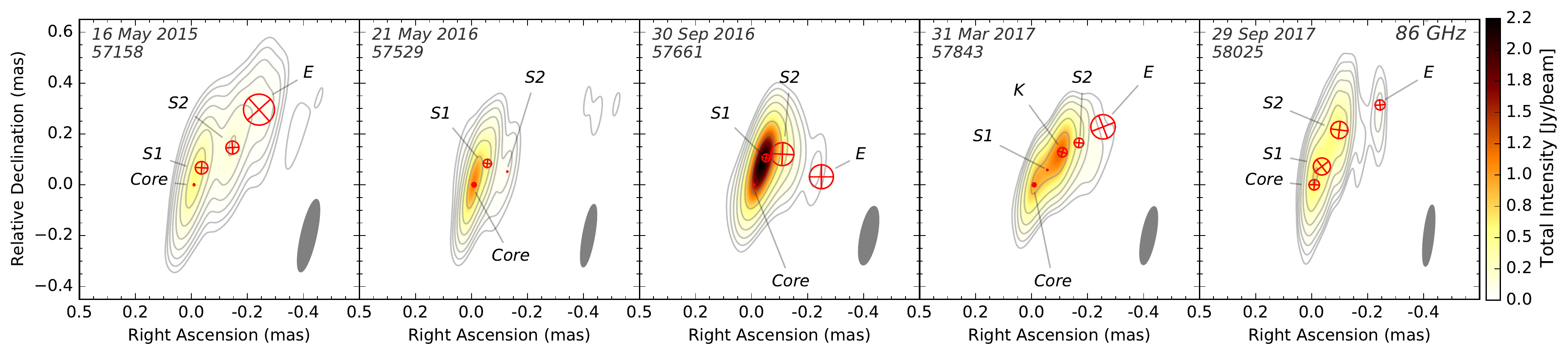}  
\end{center}
\caption{Natural weighted total intensity images for the five 86\,GHz GMVA observing epochs from 2015 (left) to 2017 (right). The MJD epoch is indicated in the top-left corner in each image. The restoring beam size, shown in the bottom-right corner in each image, is 0.29 mas $\times$ 0.06 mas (at -12.8$^{\circ}$), 0.25 mas $\times$ 0.04 mas (at -10.1$^{\circ}$), 0.23 mas $\times$ 0.06 mas (at -11.5$^{\circ}$), 0.21 mas $\times$ 0.05 mas (at -11.2$^{\circ}$), and 0.24 mas $\times$ 0.04 mas (at -6.5$^{\circ}$) from left to right, respectively. The overlaid lowest total intensity contour is at $1\%$ of the map peak (0.54, 1.11, 2.30, 2.53, and 0.68 Jy/beam from left to right, respectively), with the contours that follow a factor of two higher. The image rms noise from left to right is 0.54, 1.03, 2.71, 4.15, and 3.47 mJy/beam, respectively. The color scale represents the total intensity. The overlaid red circles represent the model-fit components.}
\label{img_gmva_images}
\end{figure*}

Since February 5, 2017, OJ\,287 has aroused considerable interest for another reason: the Very Energetic Radiation Imaging Telescope Array System (VERITAS) collaboration reported the first ever very high-energy (VHE; E$>100$\,GeV) $\gamma$-ray emission detection of OJ\,287 \citep{ATel10051}. Different theoretical models related to the mass accretion, the relativistic plasma outflows, and the particle pair-production processes have been considered for explaining the production of such VHE emission in active galactic nuclei \citep[AGNs; e.g.,][]{Rieger2018}.
We note that blazars dominate the population of the VHE $\gamma$-ray-detected AGNs; this is mainly because of the strong Doppler boosting effects produced in their relativistic jets, which are closely aligned to our line-of-sight. Currently, $93\%$ of the AGNs listed in the online TeVCat\footnote{\url{http://tevcat2.uchicago.edu/}} catalog are blazars (74 out of 80 objects). However, the OJ\,287-like blazars, the so-called low-frequency peaked blazars, with their synchrotron emission component peaking at frequencies $<10^{14}$\,Hz, represent less than $15\%$ of the whole TeVCat blazar sample. 

In the present work we investigate the source parsec-scale properties by means of high-resolution very long baseline interferometry (VLBI) observations, with the main goal of helping to characterize the physical mechanisms connected with the reported teraelectronvolt (TeV) emission activity. We make use of five 86\,GHz Global Millimeter-VLBI Array (GMVA) observations, covering the period 2015-2017, which allow us to access the innermost jet region at a resolution of $\sim50$\,$\mu$as ($\sim0.2$\,pc in linear scale), complemented by additional multifrequency radio observations from other monitoring programs.

The paper is laid out as follows: observations and methods in Sect.~\ref{sec_observations}; main results in Sect.~\ref{sec_results}; and general discussion and concluding remarks in Sect.~\ref{sec_discussion}. Throughout the paper we use a $\Lambda$ cold dark matter cosmology with $H_0 = 67.4$ km s$^{-1}$ Mpc$^{-1}$, $\Omega_m = 0.31$, and $\Omega_\Lambda=0.69$ \citep{Planck2020}.
At a redshift of $z = 0.306$ \citep{Nilsson2010}, 1\,mas corresponds to $\sim4.7$ pc.

\section{Observations and data analysis}  \label{sec_observations}
In this work we analyze five 86\,GHz GMVA observations of OJ\,287 performed from May 16, 2015, to September 29, 2017 -- modified Julian days (MJDs) 57158, 57529, 57661, 57843, and 58025 -- in total intensity emission. These observations are part of a monitoring program\footnote{\url{http://www.bu.edu/blazars/vlbi3mm/}} (PIs: A.~P.~Marscher and T.~P.~Krichbaum) of a sample of $\gamma$-ray-bright blazars and radio galaxies selected among the targets of the VLBA-BU-BLAZAR program\footnote{\url{https://www.bu.edu/blazars/VLBAproject.html}}.

\begin{figure*}
\begin{center}
\includegraphics[scale=0.45]{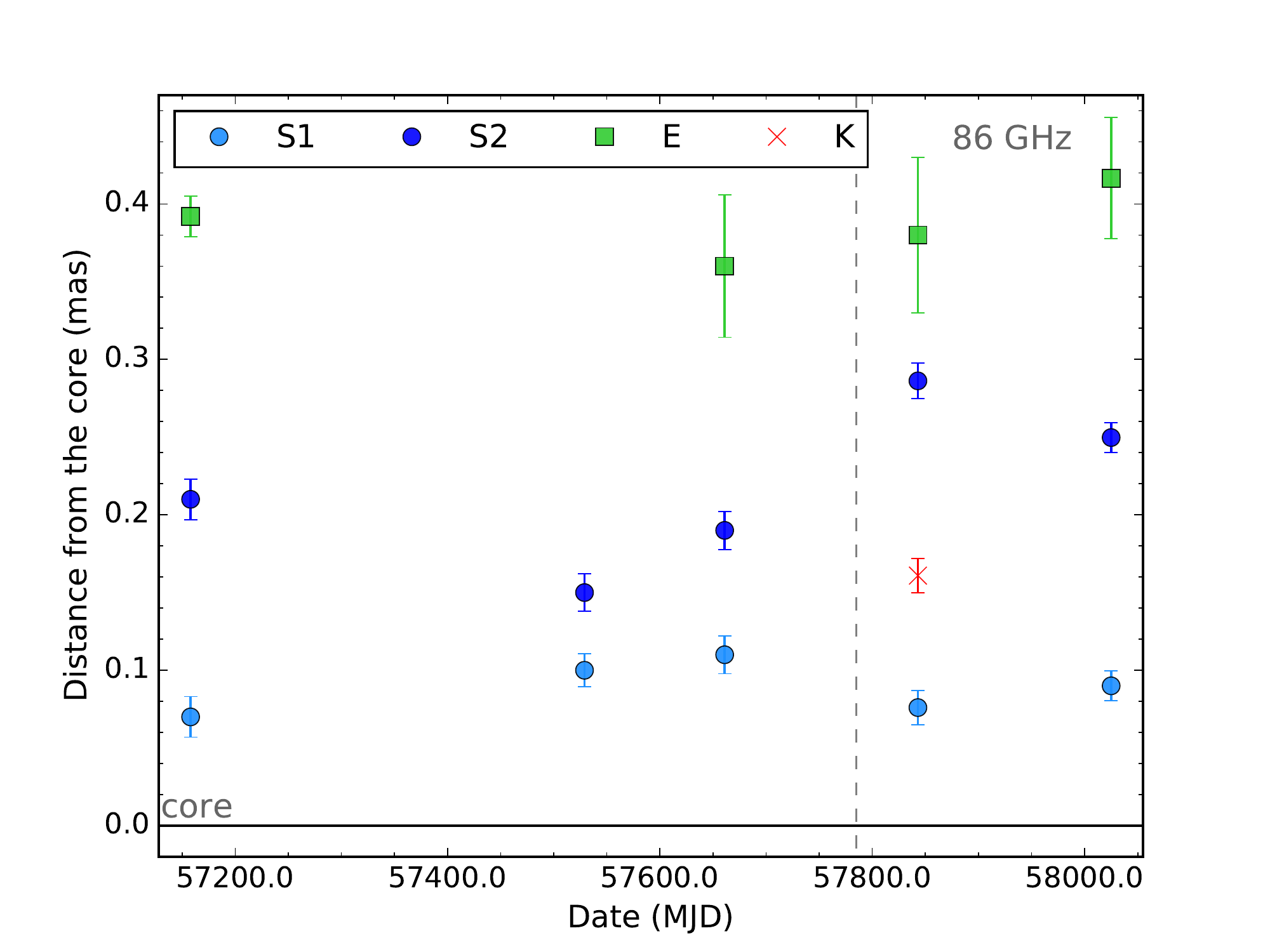}
\includegraphics[scale=0.45]{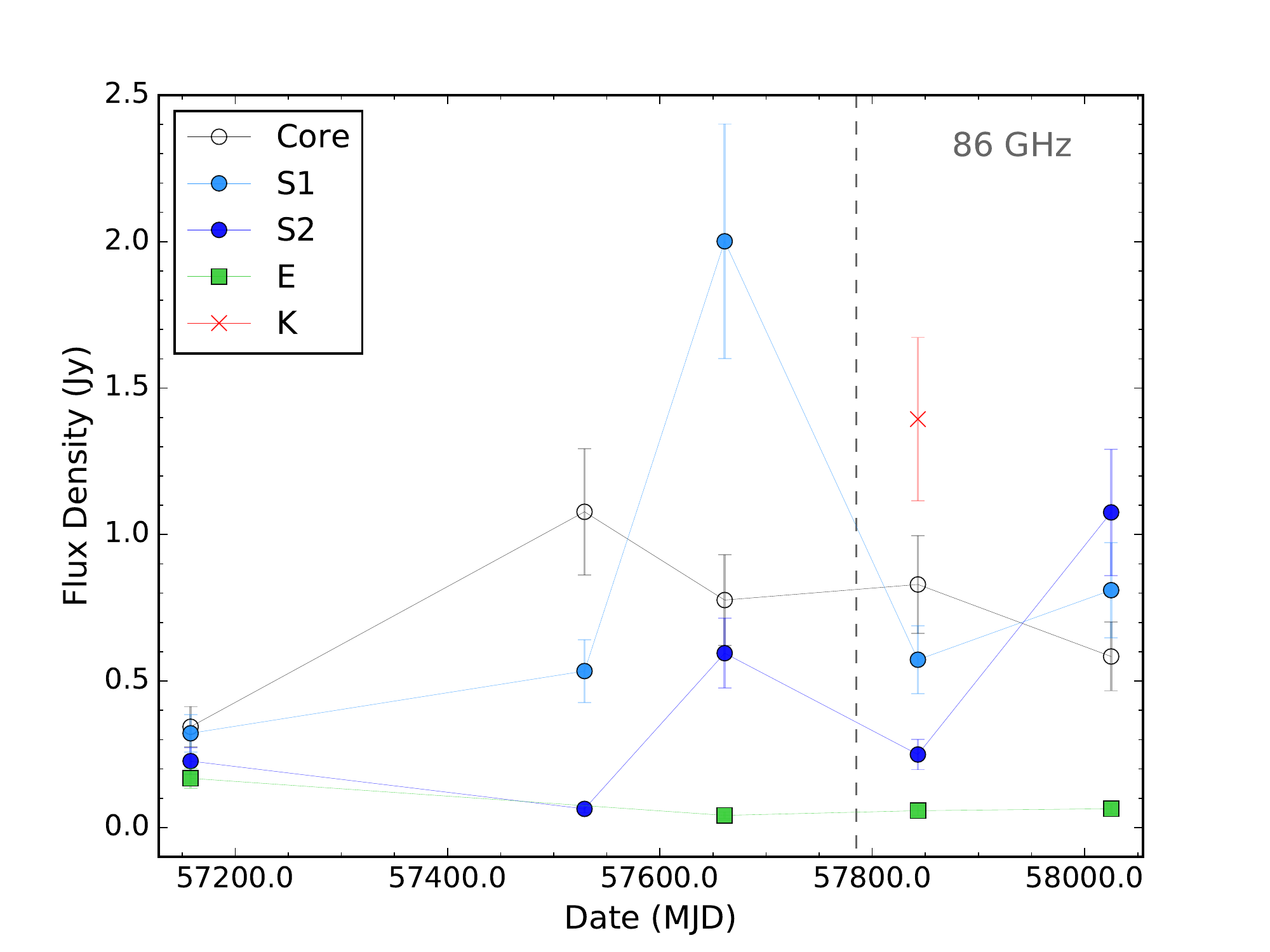}   \\
\end{center}
\caption{Time evolution of the position and flux density for each model-fit component. Left image: Time evolution of the distance from the core (represented by the horizontal solid black line) of each model-fit component. The light and dark blue circles represent components S1 and S2, respectively. The green squares and red crosses represent components E and K, respectively. The vertical dashed line represents the TeV detection (MJD 57785). Right image: Light curves for each model-fit component. The core component is represented by the empty black circles, and all the other symbols are as in the left image. The nominal total intensity emission uncertainties were calculated by considering a calibration error of $\sim20\%$ of the flux density and a statistical error equal to three times the map rms noise, but no amplitude scaling factors are provided in this work (see Sect.~\ref{sec_observations} for more details).}
\label{img_gmva_comp_motion}
\end{figure*}

The GMVA stations that took part in the different observations presented in this work are: the Robert C.~Byrd Green Bank Telescope; eight stations of the Very Long Baseline Array (VLBA): BR, FD, KP, LA, MK, NL, OV, and PT; the European stations Effelsberg, Yebes, Onsala, Mets\"aovi, and Pico Veleta; and the Korean VLBI Network array. For the a priori phase and amplitude calibration, we used the software package Astronomical Image Processing System \citep{Greisen2003}, while for the final images we used the CLEAN and self-calibration procedures in the \texttt{DIFMAP} software package  \citep{Shepherd1997}. 
More details about the calibration process of millimeter-VLBI data reduction are available in, for example, \citet{Jorstad2017}, \citet{Casadio2019}, and \citet{Kim2019}.

In each observing epoch we modeled the source brightness distribution by means of two-dimensional Gaussian circular components, using the model-fitting routine in the {\tt DIFMAP} package. The cross-identification of the model-fit components across the different epochs is based on their position, size, and flux density. The uncertainties in the position of the model-fit components were estimated by using the ratio of the size of each component to the signal-to-noise ratio \citep[as in][]{Lico2012}.

We note that, for several known reasons, significant flux losses could affect the GMVA observations \citep[e.g.,][]{Casadio2019}. The amplitude scaling factors used to correct the GMVA flux density values are usually on the order of 2.0-2.5 and are extrapolated by using close-in-time lower-frequency observations \cite[e.g.,][]{Koyama2016, Kim2019}. 
Since OJ\,287 is an extremely variable source, showing significant radio and optical emission variations on timescales of a few days \citep[e.g.,][]{Liu2017}, it is important to use observations that are as close as possible in time  to the GMVA observing epochs. 
Although several monitoring programs provide us with both single-dish and VLBI observations at different frequencies (as shown and discussed in Sect.\ref{sec_light_curves}), it was not possible to get observations that were sufficiently contemporaneous  to obtain reliable amplitude scaling factors, and we thus do not provide them here. 
The main effect of the lack of the scaling factors is that we cannot quantify the GMVA flux density evolution across the different epochs. We overcame this by using observations at multiple radio frequencies and resolutions during the same observing period (Sect. \ref{sec_light_curves}). Furthermore, the scaling factors affect neither the source structure nor the relative flux density among the different model-fit components within the same observing epoch.

\section{Results}  \label{sec_results}

\subsection{86\,GHz GMVA images and model fits}  \label{sec_GMVA_images_modelfits}
In Fig.~\ref{img_gmva_images} we report the five 86\,GHz GMVA total intensity images. The source shows a jet structure in the northwest direction, extending for $\sim 0.4$\,mas ($\sim2.0$ pc in linear scale). More details about the image beam and total intensity contours can be found in the figure caption. As is apparent from a visual inspection of the images, the source jet orientation varies across the different epochs. This is a well-known and well-investigated feature of OJ287 \citep[e.g.,][]{Cohen2017, Britzen2018, Gomez2022}. A combined multifrequency ridgeline and position angle time evolution analysis of the OJ\,287 jet, which includes all the GMVA epochs reported in this Letter (except for the last epoch, MJD 58025), was recently presented in \citet{Dey2021}.

The red circles in each sub-image represent the model-fit components. A summary of the model-fit parameters is reported in Table~\ref{table_data}. The overall source brightness distribution is well represented by four circular model-fit components, labeled Core, S1, S2, and E.
In agreement with previous works \citep[e.g.,][]{Agudo2012, Hodgson2017, Jorstad2017}, we identified the most upstream component, which is unresolved, bright, and variable,  as the core. 
The distance from the core of each model-fit component (under the assumption that the core is stationary) is reported in the left panel of Fig~\ref{img_gmva_comp_motion}.
Components S1 and S2 (light and dark blue circles in Fig~\ref{img_gmva_comp_motion}, respectively) are consistent with being stationary during the whole observing period. We note that at lower frequencies, because of the lower resolution, only a single quasi-stationary component has been detected in the images of the OJ\,287 jet since 2005, at an average distance of $0.2$\,mas from the core, as reported in several OJ\,287 kinematics studies \citep[e.g.,][]{Agudo2012, Jorstad2017, Sasada2018, Weaver2021}. The outermost jet component, labeled E, represents the more extended emission.

A new model-fit component, labeled K (red cross in Fig~\ref{img_gmva_comp_motion}), was detected in the region between S1 and S2 on MJD 57843 (March 31, 2017), about two months after the reported TeV detection. 
Although the identification of K is not obvious and straightforward, we identified the different model-fit components on MJD 57843 according to their distance from the core and flux density across the different epochs. 

\subsection{Light curves}  \label{sec_light_curves}
The light curves of the five GMVA epochs for the different model-fit components are reported in the right panel of Fig~\ref{img_gmva_comp_motion}. We find that on MJD 57661 (September 30, 2016) the overall source emission was dominated by the total intensity emission from component S1, while on MJD 57843 (March 31, 2017) it was dominated by the emission from the new component, K. During the last observing epoch, MJD 58025, component S2 became brighter than the core and S1. 

\begin{figure}
\begin{center}
\includegraphics[bb= 5 0 400 260, scale=0.64, clip]{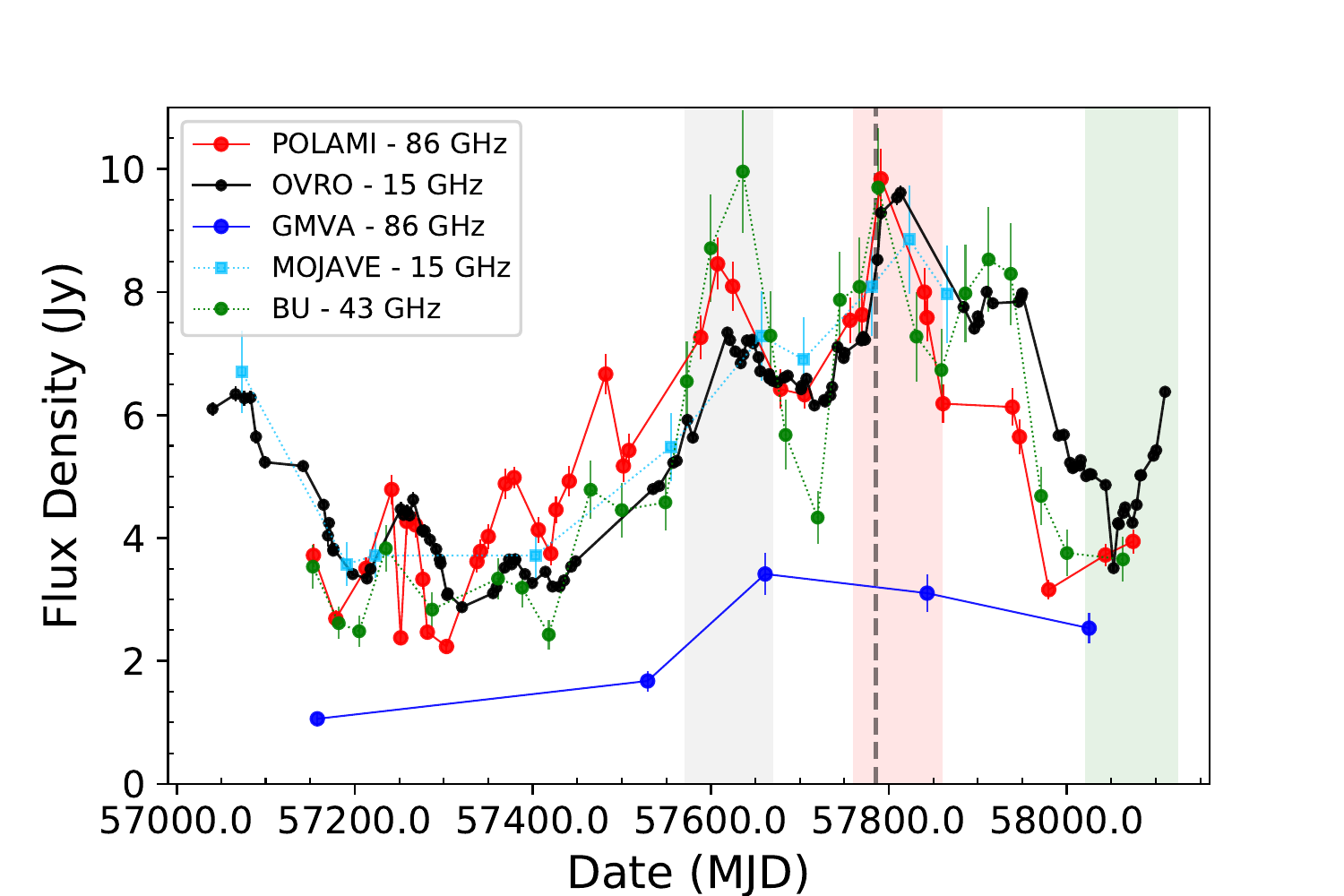}
\end{center}
\caption{Multiwavelength light curves: 15\,GHz single-dish OVRO (black circles), 86\,GHz single-dish POLAMI program (red circles), 86\,GHz GMVA (dark blue circles), 15\,GHz VLBA MOJAVE program (light blue circles), and 43\,GHz VLBA-BU-BLAZAR program (green circles) observations from 2015-2017. The vertical dashed line represents the time of the TeV detection (MJD 57785). The shaded gray, red, and green areas represent periods of enhanced emission activity (see Sects. \ref{sec_light_curves} and \ref{sec_discussion} for more details).}
\label{img_mwl_lc}
\end{figure}

To investigate the overall radio emission activity during the period 2015-2017, we produced the light curves at different radio frequencies from different monitoring programs (Fig.~\ref{img_mwl_lc}). We used: 15\,GHz single-dish observations from the Owens Valley Radio Observatory (OVRO) blazar monitoring program \citep{Richards2011}; 86\,GHz single-dish observations from the Polarimetric Monitoring of AGN at Millimetre Wavelengths (POLAMI) program\footnote{\url{http://polami.iaa.es/}} \citep{Agudo2018a, Agudo2018b, Thum2018}; 15\,GHz VLBA observations from the Monitoring of Jets in Active galactic nuclei with VLBA Experiments (MOJAVE) program\footnote{\url{https://www.physics.purdue.edu/astro/MOJAVE/}}; and 43\,GHz VLBA observations from the VLBA-BU-BLAZAR program\footnote{\url{https://www.bu.edu/blazars/VLBAproject.html}}.
As shown in Fig.~\ref{img_mwl_lc}, two periods of enhanced radio emission activity can be identified before (gray shaded area) and during (red shaded area) the VHE flaring activity, and an increasing trend is also observed toward the end of the observing window (green shaded area) at all the different radio frequencies.
We note that we have not quantified the 86\,GHz GMVA total intensity emission variability across the different epochs because we do not provide the scaling factors (see Sect.~\ref{sec_observations} for more details), and this is why in  Fig.~\ref{img_mwl_lc} the total GMVA flux density (dark blue circles) is overall underestimated. 

\section{Discussion and conclusion} \label{sec_discussion}
During the period February 1-4, 2017 (MJD 57785-57788), the blazar OJ287 was observed with VERITAS for a total exposure time of 13 hours, resulting in the first ever VHE $\gamma$-ray emission detection of the source, at $>5.7\sigma$ significance level. The source VHE detection was found to be temporally coincident with increased activity at X-ray and $\gamma$-ray energies \citep{ATel10051}, as well as at centimeter and millimeter wavelengths, as shown in Fig.\ref{img_mwl_lc}.
This event triggered several multiwavelength observing efforts and follow-ups with the primary purpose of characterizing the underlying physical mechanisms \citep[e.g.,][]{OBrien2017, Gupta2019}. 

In the present work we have investigated the innermost regions of the OJ\,287 jet, where the VHE $\gamma$ rays are thought to be produced, by means of high-resolution millimeter-VLBI observations with the GMVA. One of the main findings is that during the MJD 57158 (March 31, 2017) 86\,GHz GMVA epoch we detected a new model-fit component (K), in the region between the quasi-stationary components S1 and S2, that dominates the source total intensity emission. This is the closest-in-time GMVA epoch to the VHE event detection, separated by about two months (58 days). During the same period, an enhanced overall activity is detected at different radio frequencies (shaded red area in Fig.~\ref{img_mwl_lc}).
These two findings, a high radio emission state and the detection of a new model-fit component, could be related to the VHE event and the passage (ejection) of K through (from) the S1 quasi-stationary jet component. We note that S1, as investigated and suggested in several works \citep[e.g.,][]{Agudo2011, Sasada2018, Weaver2021}, is considered a recollimation shock.
The passage of a new component through a recollimation shock associated with an enhanced activity at high energies is a quite common event that has been observed in several AGNs \citep[e.g.,][]{Marscher2010, Jorstad2016, Casadio2019}.

The identification of K on MJD 57843 (March 31,  2017) is driven and supported by the following considerations. On MJD 57661 (September 30, 2016), when the source underwent another episode of enhanced radio emission activity (shaded gray area in Fig.~\ref{img_mwl_lc}), the overall GMVA total intensity emission was dominated by component S1, found to be more than two times brighter than the core and S2 (right image in Fig.\ref{img_gmva_comp_motion}). We interpret this finding as an indication that during that epoch the new component, K, is blended with and passing through S1 and that it emerges from the other side in the early months of 2017. Within this scenario, numerical simulations suggest that fluctuations in the position of a recollimation shock undergoing the crossing of a moving disturbance are expected \cite[e.g.,][]{Gomez1997}. We note that as K exits the S1 region and approaches S2, as predicted by simulations, both S1 and S2 show fluctuations in their positions with respect to the core (as is apparent in the left image of Fig.\ref{img_gmva_comp_motion}). In more detail, from MJD 57661 (September 30, 2016) to MJD 57843 (March 31, 2017) component S2 moves away from the core from $0.19\pm0.01$\,mas to $0.29\pm0.01$\,mas, while S1 moves toward the core region from $0.11\pm0.01$\,mas to $0.08\pm0.01$\,mas.

If we assume that component K was blended together with S1 (on average at $\sim0.1$\,mas from the core) during the VHE outburst (MJD 57785) and then observed at a distance of 0.16\,mas during the GMVA epoch MJD 57843, we obtain a propagation speed of $0.32$\,mas/year, corresponding to $\sim$4.8\,c. 
We note that more than half of the OJ\,287 knots found at 43\,GHz, from the VLBA-BU-BLAZAR monitoring program, show a proper motion between 0.3 and 0.5 mas/year \cite[e.g.,][]{Jorstad2017}.

The main findings, to be taken into account when drawing a plausible physical scenario, can be summarized as follows: (1) detection of a new jet feature between S1 and S2 in the GMVA image during the closest GMVA epoch to the TeV detection; (2) high radio emission state during the TeV flaring activity period from both single-dish and VLBI observations; (3) source total intensity emission dominated by S1 in September 2016 (prior to the TeV detection); (4) fluctuations in the positions of S1 and S2 with respect to the core as the new component, K, leaves the S1 region and approaches S2. Based on all these results, we conclude that the February 2017 TeV flaring activity in OJ\,287 could have been triggered by the passage of K through a recollimation shock represented by the quasi-stationary model-fit component S1. 
We note that, for OJ\,287, both the core and S1 regions were found to be suitable sites for particle acceleration associated with moving shocks and responsible for high energy flares \citep[e.g.,][]{Agudo2011, Hodgson2017, Sasada2018}. 
In the specific case of the 2017 OJ\,287 TeV detection, we conclude that the region where the VHE emission originates coincides with S1, located at $\sim 0.1$\,mas from the core, corresponding to a de-projected distance of $\sim10.0$ pc (by assuming a viewing angle of $2.6^{\circ}$ as reported in \citealt{Jorstad2017}).

Some additional remarks about the last GMVA observing epoch, MJD 58025 (September 29, 2017), further support the above proposed scenario. On one hand, during this last GMVA epoch component K is not detected and component E has a position and relative flux density consistent with the previous epochs. On the other hand, component S2 has a higher flux density than the core and S1, unlike during the previous epochs after 2015. These findings may indicate that during this epoch K had reached and was passing though component S2, which produced an increase in its flux density. According to the speed that we estimated for K ($0.32$\,mas/year), it is expected to have reached the region where component S2 is found by MJD 58025. 
A detailed analysis of the OJ\,287 jet kinematic properties will be presented in a dedicated and forthcoming paper based on observations from the 43\,GHz VLBA-BU-BLAZAR monitoring \citep{Weaver2021}. 
We also note that after the overall flux density decrease following the TeV flaring activity, the light curves at different radio frequencies reported in Fig.~\ref{img_mwl_lc} (shaded green area) show a new increasing trend toward the end of 2017.

\begin{table}
\begin{center}
\begin{tiny}
\caption{Summary of the model-fit parameters.}
\label{table_data}   
\setlength{\tabcolsep}{3.2pt}
\renewcommand{\arraystretch}{1.0}     
\begin{tabular}{ccccccccc}  
\hline\hline                 
Obs. date & MJD & Comp. ID & $S$\tablefootmark{a} & $\sigma_{S}$\tablefootmark{b} & r\tablefootmark{c} & $\sigma_r$\tablefootmark{d} & PA\tablefootmark{c} & FWHM\tablefootmark{e}\\
yy/mm/dd &  & & (Jy) & (Jy) & (mas) & (mas) & (deg) & (mas) \\
\hline\hline     
&&&&&&&&\\
2015/05/16   & 57158 & Core & 0.34 & 0.07 & ...  & ...  & ..... & 0.006 \\
                 &       & S1   & 0.32 & 0.06 & 0.07 & 0.01 & -22.2 & 0.049 \\ 
                 &       & S2   & 0.23 & 0.05 & 0.21 & 0.01 & -46.2 & 0.051 \\ 
                 &       & E    & 0.17 & 0.03 & 0.39 & 0.01 & -40.9 & 0.122 \\ 
2016/05/21   & 57529 & Core & 1.08 & 0.22 & ...  & ...  & ..... & 0.017 \\ 
                 &       & S1   & 0.53 & 0.11 & 0.10 & 0.01 & -32.2 & 0.033 \\ 
                 &       & S2   & 0.06 & 0.01 & 0.15 & 0.01 & -68.4 & 0.004 \\ 
2016/09/30   & 57661 & Core & 0.78 & 0.16 & ...  & ...  & ..... & 0.006 \\ 
                 &       & S1   & 2.00 & 0.40 & 0.11 & 0.01 & -23.2 & 0.030 \\ 
                 &       & S2   & 0.60 & 0.12 & 0.19 & 0.01 & -42.8 & 0.090 \\ 
                 &       & E    & 0.04 & 0.01 & 0.36 & 0.05 & -83.5 & 0.095 \\
2017/03/31   & 57843 & Core & 0.83 & 0.17 & ...  & ...  & ...   & 0.016 \\
                 &       & S1   & 0.57 & 0.12 & 0.08 & 0.01 & -41.8 & 0.005 \\
                 &       & S2   & 0.25 & 0.05 & 0.29 & 0.01 & -47.0 & 0.038 \\
                 &       & K    & 1.39 & 0.28 & 0.16 & 0.01 & -40.6 & 0.038 \\
                 &       & E    & 0.06 & 0.02 & 0.38 & 0.05 & -50.2 & 0.096 \\
2017/09/29   & 58025 & Core & 0.58 & 0.12 & ...  & ...  & ...   & 0.041 \\ 
                 &       & S1   & 0.81 & 0.16 & 0.09 & 0.01 & -22.6 & 0.067 \\ 
                 &       & S2   & 1.08 & 0.22 & 0.25 & 0.01 & -24.3 & 0.068 \\ 
                 &       & E    & 0.06 & 0.02 & 0.42 & 0.04 & -39.6 & 0.040 \\ 
\hline                                  
\end{tabular}
\end{tiny}
\tablefoot{
\begin{tiny}
\newline
\tablefoottext{a}{Flux density ($S$) in Jy;} \tablefoottext{b}{uncertainties on $S$;}
\tablefoottext{c}{$r$ and $PA$ represent the distance and position angle (from north to east) of the component in polar coordinates with respect to the core;} \tablefoottext{d}{uncertainties on $r$;}
\tablefoottext{e}{radius of the circular Gaussian components.}
\end{tiny}
}
\end{center}
\end{table}

In conclusion, with the current work we (1) identify as a possible trigger of the 2017 TeV enhanced activity the passage of a new jet feature through a recollimation shock and (2) determine that the jet region where this occurs is located at a de-projected distance of $\sim10$ pc from the core. 
These findings add a new important piece of information to a wider ongoing multiwavelength analysis that will help to constrain emission models and characterize details about the mechanism(s) producing the observed VHE emission (VERITAS collaboration et al.\,in prep.). 
A detailed multiband optical photometric and polarimetric analysis of OJ\,287, covering the period from September 2016 to December 2017, was presented in \citet{Gupta2019}, and the 86\,GHz polarization analysis with GMVA+ALMA observations from 2017 will be presented in a forthcoming publication by Zhao et al.
Furthermore, the results of this Letter are meant to complement and be part of a series of other ongoing works that feature the analysis of OJ\,287 at a finer resolution by means of 230\,GHz Event Horizon Telescope observations (G\'omez et al.\,in prep) and 22\,GHz space-VLBI RadioAstron observations (Lico et al.\,in prep).

\begin{acknowledgements} \label{aknowlwdgements}
This research has made use of data obtained with the Global Millimeter VLBI Array (GMVA), which consists of telescopes operated by the MPIfR, IRAM, Onsala, Metsahovi, Yebes, the Korean VLBI Network, the Greenland Telescope, the Green Bank Observatory and the Very Long Baseline Array (VLBA). The VLBA and the GBT are a facility of the National Science Foundation operated under cooperative agreement by Associated Universities, Inc. The data were correlated at the correlator of the MPIfR in Bonn, Germany. RL, JLG, GYZ, AF, TT, IC, IA and AA acknowledge financial support from the State Agency for Research of the Spanish MCIU through the “Center of Excellence Severo Ochoa” award for the Instituto de Astrofísica de Andalucía (SEV-2017-0709), from the Spanish Ministerio de Economía y Competitividad, and Ministerio de Ciencia e Innovaci\'on (grants AYA2016-80889-P, PID2019-108995GB-C21, PID2019-107847RB-C44), the Consejería de Economía, Conocimiento, Empresas y Universidad of the Junta de Andalucía (grant P18-FR-1769), the Consejo Superior de Investigaciones Científicas (grant 2019AEP112). 
This study makes use of 43 GHz VLBA data from the VLBA-BU Blazar Monitoring Program (BEAM-ME and VLBA-BU-BLAZAR;http://www.bu.edu/blazars/VLBAproject.html), funded by NASA through the Fermi Guest Investigator Program. The research at Boston University was supported in part by NASA Fermi Guest Investigator program grant 80NSSC20K1567. The VLBA is an instrument of the National Radio Astronomy Observatory. The National Radio Astronomy Observatory is a facility of the National Science Foundation operated by Associated Universities, Inc. The POLAMI observations were carried out at the IRAM 30m Telescope. IRAM is supported by INSU/CNRS (France), MPG (Germany) and IGN (Spain). This research has made use of data from the MOJAVE database that is maintained by the MOJAVE team \citep{Lister2018}. This research has made use of data from the OVRO 40-m monitoring program \citep{Richards2011}, supported by private funding from the California Insitute of Technology and the Max Planck Institute for Radio Astronomy, and by NASA grants NNX08AW31G, NNX11A043G, and NNX14AQ89G and NSF grants AST-0808050 and AST- 1109911.
\end{acknowledgements}

\end{document}